\documentclass[12pt,english,twoside]{article}

\usepackage[longnamesfirst]{natbib}
\usepackage{url}
\makeatletter
\def\url@leostyle{%
    \def\UrlFont{\sf}}{\def\UrlFont{\small\ttfamily}}
\makeatother
\usepackage{geometry}
\usepackage{fancyhdr}
\usepackage[font=small,labelfont=bf]{caption}

\usepackage[hidelinks,breaklinks]{hyperref}
\usepackage[hyphenbreaks]{breakurl}

\usepackage{authblk}

\usepackage{enumitem}

\usepackage{amsmath}

\usepackage[T1]{fontenc}
\usepackage[charter]{mathdesign} 

\usepackage{microtype}

\usepackage{graphicx}

\usepackage[english]{babel}

\numberwithin{equation}{section}

\usepackage{titlesec}

\titleformat{\section}[runin]
  {\normalfont\normalsize\bfseries}{\S\thesection.}{1em}{}
\titleformat{\subsection}[runin]
  {\normalfont\normalsize\bfseries}{\S\thesubsection.}{1em}{}
\titleformat{\subsubsection}[runin]
  {\normalfont\normalsize\bfseries}{\S\thesubsubsection.}{1em}{}

\defcitealias{kant1781pluhar}{CPR}
\defcitealias{kant1785}{\emph{Grounding}}  
\defcitealias{kant1790pluhar}{CJ}
\defcitealias{kant1755}{\emph{New Elucidation}}
\defcitealias{kant1763a}{\emph{Only Possible Argument}}
\defcitealias{kant1763b}{\emph{Inquiry}}
\defcitealias{kant1770}{\emph{Dissertation}}
\defcitealias{kant1783}{\emph{Prolegomena}}

\usepackage{color}
\newcommand{\blue}[1] {%
  {\color{blue} #1}%
}

\newcommand{\documentTitle}{Kantian and Neo-Kantian First Principles for
  Physical and Metaphysical Cognition}

\title{\documentTitle}
\author[a, b]{Michael E. Cuffaro}
\affil[a]{{\small University of Western Ontario, Rotman
    Institute of Philosophy}}
\affil[b]{{\small Ludwig-Maximilians-Universit\"at
    M\"unchen, Munich Center for Mathematical Philosophy}}
\date{}

\begin{document}

\maketitle

\thispagestyle{empty}

\begin{abstract}
I argue that Immanuel Kant's critical philosophy---in particular the doctrine of transcendental idealism which grounds it---is best understood as an `epistemic' or `metaphilosophical' doctrine. As such it aims to show how one may engage in the natural sciences and in metaphysics under the restriction that certain conditions are imposed on our cognition of objects. Underlying Kant's doctrine, however, is an ontological posit, of a sort, regarding the fundamental nature of our cognition. This posit, sometimes called the `discursivity thesis', while considered to be completely obvious and uncontroversial by some, has nevertheless been denied by thinkers both before and after Kant. One such thinker is Jakob Friedrich Fries, an early neo-Kantian who, despite his rejection of discursivity, also advocated for a metaphilosophical understanding of critical philosophy. As I will explain, a consequence for Fries of the denial of discursivity is a radical reconceptualisation of the method of critical philosophy; whereas this method is a priori for Kant, for Fries it is in general empirical. I discuss these issues in the context of quantum theory, and I focus in particular on the views of the physicist Niels Bohr and the Neo-Friesian philosopher Grete Hermann. I argue that Bohr's understanding of quantum mechanics can be seen as a natural extension of an orthodox Kantian viewpoint in the face of the challenges posed by quantum theory, and I compare this with the extension of Friesian philosophy that is represented by Hermann's view.\footnote{A shorter version of this paper, focusing specifically on the views of Grete Hermann, has been published as: \blue{\href{https://link.springer.com/chapter/10.1007/978-3-031-08593-2_6}{Cuffaro, Michael (2023).} ``Grete Hermann, Quantum Mechanics, and the Evolution of Kantian Philosophy.'' In Jeanne Peijnenburg \& Sander Verhaegh (eds.), \emph{Women in the History of Analytic Philosophy}. Cham: Springer. pp. 114-145.} The published version provides a sharper (and clearer) presentation of both Kant's doctrine of transcendental idealism as well as of Grete Hermann's views on quantum mechanics and is generally to be preferred. However this longer preprint contains a more extended discussion and defense of the epistemic reading of transcendental idealism in particular, as well as an extended discussion of the views of Niels Bohr.}
\end{abstract}

\section{}
\label{sec:intro}

Even before Hegel's death in 1831, the idea, traditional since the ancient Greeks, that philosophical speculation could furnish a general model for scientific thought had been waning in the Germanic world. What were seen as the excesses of the romantic and idealist movements, combined with the rapid growth and manifest success of empirical science, had brought philosophy into disrepute. Some---the so-called `vulgar materialists'---went so far as to call for philosophy to essentially identify itself with empirical science (\citealt[ch. 3]{schnadelbach1984}; \citealt[pp. 182-188]{beiser2014}).\nocite{bacciagaluppi2017}

The neo-Kantian movement of the mid-nineteenth century---itself a revival of an earlier trend of thought contemporaneous with speculative idealism and exemplified by such thinkers as Jakob Friedrich Fries \citep[ch. 1]{beiser2014}---arose partly as a reaction to these intellectual currents. Championed, at first, primarily by natural scientists such as Helmholtz \citep[p. 103]{schnadelbach1984}, the movement did not, as did speculative idealism, view it as philosophy's task to produce scientific knowledge by pure thought alone. Nor did these neo-Kantians, with the vulgar materialists, consider it philosophy's task merely to uncritically catalogue and systematise the results of empirical science. Rather, with Kant, they saw it as philosophy's essential, unavoidable, and enduring mission to enquire into the sources of our knowledge and the degree of its justification.\footnote{\emph{... sondern sie beabsichtigte nur, die Quellen unseres Wissens und den Grad seiner Berechtigung zu untersuchen, ein Gesch\"aft, welches immer der Philosophie verbleiben wird, und dem sich kein Zeitalter ungestraft wird entziehen k\"onnen} \citep[p. 5]{helmholtz1855}.} In other words they, and the philosophers such as Liebmann, Lange, and Cohen who took up their battle call \citep[]{zeman1997}, saw the proper task of philosophy as consisting in the provision of an \emph{epistemological foundation} for science. By the close of the nineteenth century, neo-Kantianism exerted a powerful influence on Germanic thought.

It was against this intellectual backdrop that many of the modern era's spectacular achievements in logic, mathematics, and physical science were made. Some of these were to eventually deal a heavy blow to the popularity of Kant's philosophy. Pure logic, Kant had argued, could never provide us with a genuine expansion of our knowledge; yet Frege's \emph{Begriffsschrift} and the later systems that drew from it seemed to provide us with tools to do just this. Euclidean geometry was held by Kant to be both a synthetic and an a priori science. Yet Hilbert was able to show that it followed analytically from a set of basic axioms, and the development and physical application (in relativity theory) of non-Euclidean geometry showed that no one particular geometry could be regarded as a priori true. Arguably worst of all for Kant, the development of quantum theory seemed to tell against according a fundamental status to the principle of cause and effect. By the middle of the twentieth century it was widely held that Kant's philosophy had been definitively refuted.\footnote{See, for instance, Carnap's summary of the prevailing attitude toward Kant in \citet[p. vi]{reichenbach1958}.}

Yet the truth is more subtle than this. Many of the late-nineteenth and early twentieth century thinkers whose work had contributed to the demise in popularity of Kantian philosophy were, despite their divergences from him, substantially influenced by Kant. Frege, for example, is at pains to call attention to his agreement with Kant, which he claims far exceeds the extent of his disagreement \citep[\textsection 89]{frege1980}.\footnote{For more on the parallels between Kant and Frege, see \citet[]{cuffaro2012a, merrick2006}.} In the epigraph to his seminal work on geometry, Hilbert expresses the affinity of his thought with Kant's by invoking the latter in support of the spirit of his investigations \citep[p. 1]{hilbert1902}.\footnote{For more on the Kantian aspects of Hilbert's thought, see \citet[]{kitcher1976}.} Reichenbach's conception of the relativised a priori, the conventionalisms of Poincar\'e, Schlick, and Carnap, the pragmatisms of C. I. Lewis and others, are best characterised, not as radical rejections of Kant's philosophical framework, but rather as attempts to re-explicate the basic Kantian idea that our theoretical frameworks include an element---what Kant had (mistakenly, according to these thinkers) called the synthetic a priori---that is conceptually and epistemologically privileged in some sense. Viewed as a research program \citep[cf.][]{bitbol2017}, one may say that Kant's transcendental approach to philosophy continued, and continued to evolve, albeit along multiple independent pathways, well into the last century. It is only with the rise of the Quinean holistic conception of science that these ideas are rejected in their totality.\footnote{I am of course not claiming that all (or even any) of the thinkers mentioned in this paragraph would have called themselves Kantians; in particular Schlick and (the later) Carnap certainly would not have done so, and nor should we. I nevertheless do think that their views as well as the others I have mentioned can be seen as continuous with (in the sense of evolving continuously out of) a research programme that was begun by Kant. For more on these topics, see \citet[]{coffa1991, disalle2002, friedman2009, howard1994, murphey2005}.}

The development of quantum theory in the early part of the last century posed a particularly strong challenge to the Kantian philosophical viewpoint. Owing to the indeterminacy intrinsic to the theory, a common opinion expressed at the time was that this represents a definitive refutation of Kant's philosophy insofar as it shows that Kant's ascription of a priori status to the principle of causality cannot be correct. As with other developments in the mathematical and natural sciences during the period, however, the relationship between quantum theory and Kant's theoretical philosophy is far more rich and interesting than this. There has been a flowering of scholarship in recent years exploring the relationship between Kantian philosophy and quantum mechanics, and especially between Kant and the influential physicist Niels Bohr. Commentators such as \citet{kaiser1992}, \citet{chevalley1994}, \citet[]{pringe2009}, \citet{cuffaro2010}, \citet[]{bachtold2017}, \citet{bitbol2017}, \citet{kauark-leite2017}, and others have over the years demonstrated that Bohr's views on quantum mechanics are broadly Kantian in the sense that they can be motivated from considerations that arise naturally from within Kant's philosophical framework; further that Bohr's views remain broadly compatible with---and indeed are very much in the spirit of---a Kantian worldview adapted to address the situation presented to us by quantum theory.

Most of this literature is aimed at clarifying the philosophical viewpoint of Bohr, or of related interpretations of quantum mechanics such as that of Werner Heisenberg \citep[]{camilleri2005}. One of the contributions of this paper, in contrast, will be to turn an eye back toward Kantian philosophy. In the sequel it will be my contention that considering Kantian philosophy in the light of the challenges posed by quantum mechanics illuminates the significance of several elements of the Kantian philosophical framework, foremost among which are Kant's synthetic a priori principles (including the principle of causality) as well as his core doctrine of transcendental idealism.

Transcendental idealism is made up of two claims: first, that space and time are the necessary subjective forms of all appearances for us; second, that they do not attach to things as they are in themselves. With respect to the latter claim, there is dispute within Kant scholarship over whether to interpret it ontologically (see especially \citealt[]{guyer1987}), as a claim about how things as they exist independently of us in fact are not, or merely epistemically (see especially \citealt[]{allison2004}), as a claim about what we are licensed to attach to our conception of things as they are in themselves---in other words what we are in a position to determinately assert of them---given our particular epistemic limitations.

Allison's interpretation of Kant, at least in this general sense, is in my view the correct one. However my goal here is not to somehow refute the ontological interpretation of Kant's views. My goal, rather, is (in part) to address the charge against the epistemic reading of Kant that it amounts to a trivialisation of Kant's views grounded on no firmer a basis than bare stipulation. As such it does not, it has been argued, sufficiently capture the real ontological significance that Kant seems to accord to the transcendental distinction between appearances and things in themselves. And it does not, it has been argued, sufficiently capture the sense of `epistemic loss' one feels as a result of this inability to know things as they are in themselves.

Some stipulations are better than others, however, and in the sequel I will argue that Kant's doctrine of transcendental idealism, though it should be understood `merely epistemically', is nevertheless profoundly significant. As I will show, Kant's views on the matter are rooted in his long-running pre-critical struggle to discover the first principles for metaphysical cognition (i.e. cognition of things as they actually are in themselves). Moreover, underlying Kant's doctrine is an ontological posit of a sort; however this posit on which transcendental idealism rests---the so-called `discursivity thesis'---is not an ontological posit which pertains to the mind-independent world but is rather a posit concerning the nature of our cognition. And far from being, as some have claimed \citep[cf.][Ch. 1]{strawson1966}, a self-evident truth, discursivity is a posit which has been denied---and with interesting consequences, as we will see.

The question of the epistemic constraints and preconditions for objective cognition---the question at the heart of the doctrine of transcendental idealism interpreted epistemically---is arguably more pressing in quantum mechanics than in any other area of physics \citep[see][\S 6]{pitowsky1994}. And it is precisely the question which occupied physicists, such as Bohr, in their efforts to interpret quantum mechanics in the initial stages of the theory's development. It is also the question which during this period occupied Grete Hermann, a philosopher-mathematician who studied under Emmy Noether as well as the Friesian philosopher Leonard Nelson. Fries, like Kant, saw transcendental idealism as an epistemic doctrine. Fries, however, denied the discursivity thesis. And below we will see how this difference between the orthodox Kantian and Friesian neo-Kantian understandings of the nature of cognition manifests itself in a subtle but important difference in the lessons each inevitably draws for natural philosophy from quantum theory. I will argue that on an orthodox Kantian viewpoint---for which the view of Niels Bohr may be taken as representative---the point emphasised is that our fundamental (and necessary) forms of knowledge are ultimately idealisations and as such are simultaneously applicable only within a limited domain. We may nevertheless---because of this very limitation---continue to employ these `classical' idealisations outside of this domain in order to convey information to one another in a qualified way; i.e. with the understanding that the information conveyed in this way does not in general constitute an objective description of the phenomena under investigation. For Hermann---whom we may take to represent the Friesian viewpoint---there is, in addition to this, a deeper lesson as well: Quantum mechanics shows us that physical knowledge is fundamentally `split'. For Hermann, that is, there is a notion of classical objectivity that is applicable to quantum phenomena, however it is a form of objectivity which is valid only from within a particular perspective. Hermann's Friesian doctrine moreover provides an account of the relationship between physics and the a priori forms of knowledge which makes comprehensible how the latter may evolve and undergo revision as a consequence of developments in the former.

From here on the paper will proceed as follows. In \S\ref{sec:negalt} I will introduce Kant's doctrine of transcendental idealism and in \S\ref{sec:firstprinc} I will show how it is motivated, for Kant, by his struggle to provide first principles for metaphysical cognition in the period leading up to the publication of the first Critique. In \S\ref{sec:synthap} I will discuss how Kant's search for the first principles of metaphysical cognition was transformed, during the critical period, into the search for the first principles of synthetic a priori cognition, focusing in particular on Kant's conception of the principle of causality. In \S\ref{sec:quantmech} I will discuss the challenges faced by Kant's view which arise from the emergence of quantum theory. Then in \S\S\ref{sec:conceptind}--\ref{sec:schlick} I will show how the Kantian principles, strictly speaking, remain valid within quantum mechanics, and that (as Bohr argues) even outside of their domain of simultaneous applicability they may be used to convey information regarding the objects of our investigation in a qualified way. In \S\S\ref{sec:hermann}--\ref{sec:herm-transcideal} I then consider the views of the Friesian neo-Kantian philosopher Grete Hermann and discuss her understanding of the general situation regarding Kantian critical philosophy \emph{vis \'a vis} quantum theory. I conclude with a brief discussion of the relative merits of the orthodox Kantian and Friesian viewpoints with respect to the way they account for the challenges posed to critical philosophy by quantum theory and the way they envision critical philosophy's proper response to this challenge.

\section{}
\label{sec:negalt}

In the \emph{Critique of Pure Reason}, Kant explains that objective experience comprises two aspects: \emph{intuition}---the `this', the `that' of experience---and the \emph{concepts} whereby one synthesises the manifold of intuition. Concepts belong to the faculty of \emph{understanding}, which we will discuss in \S \ref{sec:synthap}. Intuition is mediated by the faculty of \emph{sensibility}: our mind's capacity to be affected by objects (A19/B33).\footnote{References to the \emph{Critique of Pure Reason}, or `first Critique', will be to the Pluhar translation \citeyearpar{kant1781pluhar}. Page numbers for Kant's works cited in this essay (with the exception of the first Critique) will be as they are in the standard German edition of Kant's works. In the case of the first Critique, page numbers are as in the first (1781) and second (1787) edition, where ``A'' denotes the first and ``B'' the second, as usual.} The effect on sensibility of some object is called the \emph{sensation} of it, and with sensation we associate the \emph{empirical} aspect of our intuition.

We call ``[t]he undetermined object of an empirical intuition'' (A20/B34) an \emph{appearance}. A shape against the far wall in a dark room, for instance, which only after some scrutiny is determined to be a chair, is before this determination merely the appearance of something indeterminate.\footnote{For a more thorough discussion of this point, see \citet[p. 110-111]{harper1984}.} There are two aspects to an appearance. First, there is its \emph{matter}; i.e. \emph{what} we sense in it. Second, there is an appearance's \emph{form}, i.e. that which allows the manifold corresponding to the appearance to be ordered in certain relations. Appearance has two characteristic forms: space, associated with outer appearances, and time, associated with both inner and outer appearances. As forms of appearances, they are the \emph{formal conditions} for appearances, in virtue of which they are known a priori as necessary relations according to which sensations must be ordered for subjects like us (A20/B34, A26/B42, A33/B49--50). They are also called `pure' in virtue of not in themselves containing anything belonging to the matter of sensation (A20/B34).

With respect to the intuitions of space and time, Kant's doctrine of \emph{transcendental idealism} puts forward two theses. The thesis just asserted, that space and time are necessary subjective conditions for appearances, is sometimes called the `subjectivity thesis'. Kant also makes the following claim:

\begin{quote}
Space represents no property whatever of any things in themselves, nor does it represent things in themselves in their relation to one another. That is, space represents no determination of such things ... that adheres to objects themselves and that would remain even if we abstracted from all subjective conditions of intuition. (A26/B42).
\end{quote}

This, and the parallel claim for time (A33/B49), are together sometimes referred to as the `nonspatiotemporality thesis'. Of the subjectivity and nonspatiotemporality theses, it is nonspatiotemporality which is more controversial. For on one interpretation of Kant's argument for nonspatiotemporality, Kant concludes that space and time cannot be determinations of things in themselves \emph{because} they are necessary subjective conditions of intuition. Assuming this reading of Kant is correct, such a conclusion would only follow if subjectivity and spatiotemporality were mutually exclusive options. But some have argued that there is no reason to think so. In particular some have pointed out that the very fact that space and time are necessary forms of appearances may be taken as good evidence that things in themselves are spatiotemporal.

This objection, that Kant has `neglected an alternative', is most often associated with Trendelenburg \citep[p. 107]{gardner1999}, but the objection was also made much earlier by others, for instance by Feder \citep[p. 140]{sassen2000}. Among more recent commentators, Guyer puts it particularly forcefully:

\begin{quote}
Transcendental idealism is not a skeptical reminder that we \emph{cannot be sure} that things as they are in themselves \emph{are} also as we represent them to be; it is a harshly dogmatic insistence that we \emph{can be quite sure} that things as they are in themselves \emph{cannot be} as we represent them to be. \citep[p. 333]{guyer1987}.
\end{quote}

Guyer is actually slightly more charitable to Kant than this quote suggests, for he argues \citeyearpar[p. 366]{guyer1987} that Kant does in fact provide an (ultimately unsuccessful) argument for the exclusion of Trendelenburg's alternative. We will pass over the details of Guyer's analysis of the deficiency in Kant's argument here, except to say that Guyer's criticism---and the neglected alternative objection more generally---presupposes an ontological reading of transcendental idealism that was not Kant's.

Kant's transcendental idealism does not in fact amount to a metaphysical claim regarding the nature of things in themselves as they exist apart from us. It is best interpreted, rather, as an `epistemic' doctrine \citep[ch. 1]{allison2004}. That is to say, on Kant's view, implicit in the concept of objective cognition are the subjective---i.e. epistemic---conditions under which an object is representable to us. In particular, on Kant's view our cognition is discursive in the sense that, as we saw earlier, we possess both a sensible and a conceptual faculty; the former being that through which our intuitions of external objects are mediated, and the latter being the faculty through which we spontaneously act upon these intuitions in order to subsume them under general concepts. Discursivity amounts to the claim that we require \emph{both} in order for cognition to be possible. As Kant puts it:

\begin{quote}
Neither of these properties is to be preferred to the other. Without sensibility no object would be given to us; and without understanding no object would be thought. Thoughts without content are empty; intuitions without concepts are blind. Hence it is just as necessary that we make our concepts sensible (i.e. that we add the object to them in intuition) as it is necessary that we make our intuitions understandable (i.e., that we bring them under concepts). Moreover, this capacity and this ability cannot exchange their functions. The understanding cannot intuit anything, and the senses cannot think anything. Only from their union can cognition arise. (A51/B75--76).
\end{quote}

From the fact that cognition requires the union of sensibility and understanding, however, it does not follow that we cannot think about objects apart from the conditions pertaining to our sensibility. We may do so by abstracting away from these conditions and in this way we may form the concept of an object `regarded as it is in itself'. But since cognition, given that it is discursive, requires the contributions of both sensibility and understanding, it follows that the concept that we form in this way of a thing regarded as it is in itself is not cognisable. Thus interpreted, transcendental idealism amounts to a claim about our concept of a thing as it is in itself, and in that sense we may call this a `conceptual' doctrine: It says simply that our concept of a thing regarded as it is in itself is not spatiotemporal because in forming it we abstract away from its spatiotemporal attributes.

Note, however, that it does not follow from this understanding of transcendental idealism that all talk of things regarded as they are in themselves should be considered empty or meaningless \citep[cf.][pp. 28, 32]{ameriks2003}. What transcendental idealism proscribes is merely the assumption that things which exist independently of us can be attributed spatiotemporal characteristics either in themselves or in relation to one another \citep[cf.][p. 24, 36]{allison2004}. But this does not entail that the concept of a thing regarded as it is in itself is without significance \citep[cf.][p. 56]{allison2004}, nor that it is ontologically less (or more) privileged than the concept of a thing regarded under the conditions according to which it can be cognised by us \citep[p. 47]{allison2004}.

The idea expressed by Kant's discursivity thesis---that cognition requires the contributions of both sensibility and understanding and that therefore there can be no standpoint-independent perspective on the objects of cognition (cf. \citealt[\S 4.3.3]{stang2016})---is central to this understanding of transcendental idealism. \citet[ch. 2]{allison2004} calls this the `anthropocentric' paradigm of cognition and as such it may be contrasted with the `theocentric' paradigm. According to the latter, true cognition is standpoint-independent in an absolute sense. That is, to know something on the theocentric paradigm is to know it absolutely independently of the way it is considered. But according to Kant this is impossible for finite cognisers such as ourselves.

Understood in this way, both transcendental idealism and the `transcendental realism' associated with the theocentric paradigm may be thought of as expressing norms or standards by which to evaluate cognition, and in this sense can be thought of as `metaphilosophical' doctrines \citep[p. 35]{allison2004}. Transcendental idealism is a metaphilosophical doctrine according to which we judge philosophical assertions \emph{vis \'a vis} the kind of cognition that is achievable \emph{for us}, and Kant's critical project more generally constitutes a reconceptualisation of philosophy in light of this aim.

Of course, this way of interpreting Kant's doctrine is not without its critics. Scholarly criticisms of the epistemic\footnote{In the sequel I will (following Allison) be using the terms `epistemic interpretation', `conceptual interpretation', and `metaphilosophical interpretation' as synonymous unless I indicate otherwise.} reading include the following: that it is trivial or tautologous, that despite the name it smuggles in a substantive ontological position (regarding things in themselves) implicitly, that the epistemic version of the argument for nonspatiotemporality is no less fallacious than the ontological version, and that it is merely anodyne (where this word is used \emph{ad hominem} as a term of abuse) and does not do justice to the important ontological issues that mattered for Kant. In the next two sections I will be focusing on the last objection listed. To my mind this is, despite the unfortunate \emph{ad hominem} character of the arguments often given in support of it, both the most important of the objections listed as well as one which has been insufficiently addressed in Kant scholarship to date.\footnote{Allison's \citeyearpar[pp. 9-11]{allison2004} response to Langton, for example, only briefly addresses this `motivational' objection in connection with the charge of triviality. \citet[\S 4.4]{stang2016} mentions this objection as a way of motivating Langton's own metaphysical interpretation of transcendental idealism, but does not consider, as I do here, the question of how it might be answered from \emph{within} an epistemic interpretation.} (A survey of the other mentioned objections to the epistemic interpretation, along with some responses, is given in \citealt[\S 4.3]{stang2016}.) Rae Langton expresses the objection thus:

\begin{quote}
Allison's is an ingenious and attractive solution to an old an ugly problem. But I would like to suggest that there are reasons for suspicion. I have an ulterior motive. I have up my sleeve a solution that is, though less ingenious, more attractive ... Allison's approach makes it analytic that we have no knowledge of things in themselves. To consider things in themselves is simply to consider things in abstraction from the conditions of our knowledge: $K_3$ [that we can have no knowledge of things in themselves] has become the tautological $A_3$ [that things considered in abstraction from their relation to our sensibility are things considered in abstraction from their relation to our sensibility]. From one point of view this is an advantage, but from another it is a grave defect, for it fails to do justice to an aspect of Kant that ought not to be ignored. What I have in mind is not exactly a Kantian philosophical thesis ... Rather, it is a Kantian attitude to these philosophical theses, and in particular [$K_3$]. When Kant tells us that we have no knowledge of things in themselves, he thinks he is telling us something new and important. The truth of $K_3$ is a major philosophical discovery. Moreover, it is not just a discovery with a definite, non-trivial content. It is a depressing discovery. Kant thinks we are missing out on something in not knowing things as they are in themselves. Kant speaks of our yearning for something more, he speaks of doomed aspirations, he speaks of `our inextinguishable desire to find firm footing somewhere beyond the bounds of experience' (A796/B824). It is not easy to see how this inextinguishable desire could be for the falsity of $A_3$. \citeyearpar[p. 10]{langton1998}.\footnote{Cf. \citet[p. 334]{ameriks1992}.}
\end{quote}

I believe this concern is a legitimate one in the sense that any interpretation of Kant should account for the attitude of Kant's that is alluded to here. But in the sequel I will argue that ascribing this attitude to Kant is not at all inconsistent with an epistemic reading of transcendental idealism. Indeed, for Kant this metaphilosophical doctrine is motivated by this very attitude. For, as I will argue below, it was through his long-running pre-critical struggle to discover the first principles for metaphysical cognition---cognition of things as they exist in themselves in a mind-independent sense---that his critical doctrine of transcendental idealism was born, and with it a new set of first principles pertaining to \emph{synthetic a priori} cognition.\footnote{Kant's `pre-critical' period refers to the period of Kant's philosophy up until the publication of the \emph{Critique of Pure Reason} in 1781.} We will also see that Kant's metaphilosophical doctrine does rest on an ontological posit of a sort; only it---namely the discursivity thesis---is not an ontological posit which pertains to the mind-independent world but rather to the nature of our cognition. Later, in \S\ref{sec:fries-nelson}, we will see that this is not the only ontological posit from which to develop an anthropocentric standard of cognition.

\section{}
\label{sec:firstprinc}

Kant in fact acknowledged a version of the `neglected alternative' objection almost a decade before the publication of the first Critique. In a famous letter to Marcus Herz of February 21, 1772,\footnote{The letter is famous because it marks the beginning of Kant's so-called `silent decade' and poses many of the questions that he will later take up in the 1781 Critique.} addressing Lambert's objection to his conception of time,\footnote{Compare also the ``Elucidation'' to the Aesthetic, A36-37.} Kant wrote:

\begin{quote}
[Lambert objects that changes] are possible only on the assumption of time; therefore time is something real ... Then I asked myself: Why does one not accept the following parallel argument? Bodies are real (according to the testimony of outer sense). Now, bodies are possible only under the condition of space; therefore space is something objective and real that inheres in the things themselves. The reason lies in the fact that it is obvious, in regard to outer things, that one cannot infer the reality of the object from the reality of the representation ... \citep[p. 75]{zweig1967}.
\end{quote}

This is the same objection which Guyer puts as follows: ``Why doesn't the indispensable role of space and time in our experience prove the transcendental realism rather than idealism of space and time themselves?'' \citep[p. 349]{guyer1987}. Kant's answer, that it is ``obvious'' that such an inference is invalid, seems unsatisfactory. But before dismissing this as mere dogmatism we should attempt to understand exactly why Kant thinks it is obvious. We can understand Kant's position if we place his 1772 letter in its proper context. Kant's preoccupation, as early as 1755, was with the question of how one might make metaphysics legitimate. For Kant, although the aim of metaphysics is purportedly rational certainty with regards to its propositions, in practice its methods are often arbitrary and hypothetical in character. His (preparatory) project at this stage is thus to provide a principled methodology or \emph{epistemic grounding} for metaphysics, now to be thought of as a \emph{science}.

Thus, in the \emph{New Elucidation} of 1755, Kant declares that his intention is to ``establish two new principles of metaphysical cognition'' \citepalias[1:387]{kant1755}. ``By their means'', he tells us, ``you may acquire no inconsiderable power in the realm of truths.'' \citepalias[1:416]{kant1755}. In his \emph{Only Possible Argument} of 1763, Kant declares that he will supply---not an actual demonstration---but the only sure way in which a demonstration of God's existence may proceed: ``What I am furnishing here is the materials for constructing a building ...'' \citepalias[2:66]{kant1763a}. In his `Prize Essay' of the same year, Kant outlines the rules by which metaphysics must proceed in its investigations and ``by which alone the highest possible degree of metaphysical certainty can be attained'' \citepalias[2:285]{kant1763b}. Within metaphysics, Kant tells us, ``One's chief concern will be to arrive only at judgements about the object which are true and completely certain'' (ibid.). Rationally certain cognition with regard to the propositions of metaphysics (traditionally construed) is the goal which by the first Critique Kant will ultimately reject as unachievable. Yet at this stage, Kant is still hopeful that metaphysically certain cognition is possible if only one can purify metaphysics' methods.

Kant's solution to the problem was to take its most mature form in his \emph{Inaugural Dissertation} of 1770. Here we are told that we will achieve our goal if we first identify and abstract from the form and principles of \emph{sensible} cognition. In this way we will elucidate the form and principles of \emph{intellectual} cognition (the proper concern of metaphysics). Intellectual (or `rational') cognition is cognition which transcends the limitations imposed by our sensibility. It is subject only to the laws of the understanding: ``whatever cognition is exempt from such subjective conditions relates only to the object. It is thus clear that things which are thought sensitively are representations of things \emph{as they appear}, while things which are intellectual are representations of things \emph{as they are}.'' \citepalias[2:392]{kant1770}.

In order to obtain intellectual cognition, one must abstract from all conditions related to sensibility. This includes even the form and principles of sensible cognition. Thus, of space and time, Kant tells us ``that these notions are not \emph{rational} at all, and that they are not \emph{objective} ideas of any connection, but that they are appearances, and that, while they do, indeed, bear witness to some common principle constituting a universal connection, they do not expose it to view.'' \citepalias[2:391]{kant1770}. Note that Kant is here using the term `objective' differently from the way he will later use it in the first Critique. In the Dissertation, `objective' is meant in the sense of pertaining to the noumenal, or mind-independent, world; in other words, Kant is operating there with a theocentric standard for objective cognition. But in the Critique, as we will see shortly, Kant is not hesitant to grant `objective' status to representations that are dependent on the faculty of sensibility \emph{as such}, so long as they are not dependent on any particular sense impressions.

In any case, and again anticipating the neglected alternative objection, Kant tells us that the forms of sensibility must be excluded even though they seem to provide good evidence for some underlying analogous connection between things as they are in themselves: ``the \emph{form} of the same representation is undoubtedly evidence of a certain reference or relation in what is sensed, though properly speaking it is not an outline or any kind of schema of the object ...'' \citepalias[2:393]{kant1770}. ``[E]mpirical concepts do not,'' Kant tells us, ``in virtue of being raised to greater universality, become intellectual in the \emph{real sense}, nor do they pass beyond the species of sensitive cognition; no matter how high they ascend by abstracting, they always remain sensitive.'' \citepalias[2:394]{kant1770}.

To clarify: the fact that spatiotemporal relations necessarily attach to sensible concepts seems to provide some warrant or evidence for also attaching these (or analogous) determinations to the concept of the thing in itself considered apart from the conditions of our sensibility. Nevertheless we certainly \emph{cannot} say that it follows necessarily from the former that things as they are in themselves are related to each other in any particular way, let alone spatiotemporally. For no matter how high we abstract empirical concepts, they always refer to what is given to us via sensibility.\footnote{Kant writes, at 2:394, ``a concept of the understanding \emph{abstracts} from everything sensitive, but it is \emph{not abstracted} from what is sensitive.'' This is explained in the sentence immediately preceding: ``The former expression indicates that in a certain concept we should not attend to the other things which are connected with it in some way or other, while the latter expression indicates that it would be given only concretely, and only in such a way that it is separated from the things which are joined to it.'' He goes on: ``For this reason, it is more advisable to call concepts of the understanding `pure ideas', and concepts which are only given empirically `abstract concepts'.''} We are not warranted, therefore, to attach these attributes to our conception of the thing considered as it is in itself.

It may be helpful to consider, as an analogy, mathematics. In mathematics one refrains from attaching a certain property to a concept even when this property holds for some particular instances of the concept. Instead, we include in our concept only that which can determinately be affirmed of the concept in general. For instance, we say that matrix multiplication is non-commutative since it is not the case in general that for two matrices, $A$ and $B$, that $AB = BA$, even though, for some particular matrices, $AB$ does in fact equal $BA$. It is similar with Kant's conception of the thing in itself, not in the sense that we can attach spatiotemporal attributes to some but not all particular things in themselves (whatever that would mean), but in the sense that our concept of the thing in itself, for Kant, can only be composed of properties we are in a position to determinately affirm of it as such. But if our concept of the thing in itself as such is composed only of what we can determinately affirm \emph{of it}, then spatiotemporal attributes---the conditions pertaining to our sensibility---certainly may not be attached to this concept.

In fact the difference between sensible and intelligible concepts is a difference in kind, not in degree \citep[cf.][p. 17]{allison2004}, for space and time attach necessarily to the objects of sensible cognition, but must not be attached (in the sense explicated in the previous paragraph) to the objects of intelligible cognition \citepalias[2:394]{kant1770}.\footnote{Compare this also with A43/B60: ``Even if we could bring this intuition of ours to the highest degree of distinctness, that would still not get us closer to the character of objects in themselves. For what we would cognize, and cognize completely, would still be only our way of intuiting, i.e., our sensibility; and we would always cognize it only under the conditions attaching to the subject originally: space and time.''} And since the object of metaphysics for the Kant of the Dissertation is to attain intelligible cognition, anything pertaining to sensible cognition---no matter to how high a degree of abstraction it is raised---must be excluded. Thus in the Dissertation Kant offers to metaphysicians the ``principle of reduction'' \citepalias[2:413]{kant1770}, which asserts that any concept of the understanding to which one predicates anything belonging to space and time must not be asserted objectively (in the sense of the Dissertation), i.e., asserted as having objective validity independently of all actual---or possible---experience of it. One might object that the bar that Kant sets for objective cognition here is high indeed, but one should keep in mind Kant's project and goals: the attainment of metaphysically certain cognition. Given these goals, Kant's standards are in my view appropriate, or at any rate not out of step with his contemporaries.

In any case, once we have purged our concepts of spatiotemporal elements, the question then becomes one of the formal principle of the intelligible world. In particular, the question is ``to explain how it is possible \emph{that a plurality of substances should be in mutual interaction with each other}, and in this way belong to the same whole, which is called a world'' \citepalias[2:407]{kant1770}. The principle that is left to elucidate, that is, is the principle of causality (in a generalised sense) or what Kant elsewhere calls the principle of the determining ground.\footnote{Cf. \citetalias[\textsection 2, Proposition IV,   1:391-393]{kant1755}.} We need not discuss Kant's more detailed comments on this principle here, but it is worth noting that, just as in his earlier works, Kant does not actually provide the needed elucidation. Rather, his goal in the Dissertation is a methodological one; namely to offer this principle and the principle of reduction as the tools with which one can begin to tackle the problems of metaphysics.\footnote{Kant does offer a sketch of what one must do in order to provide a demonstration of the community of substances: ``Granted that the inference from a given world to the unique cause of all its parts is valid, then, if, conversely, the argument proceeded in the same way from a given cause, which was common to all the parts, to the connection between them and, thus, to the form of the world (although I confess that this conclusion does not seem as clear to me), then the fundamental connection of substances would not be contingent but necessary, for all the substances are \emph{sustained by a common principle}.'' \citepalias[2:409]{kant1770}.}

\section{}
\label{sec:synthap}

By 1781 Kant had, by his own account through reading Hume, been awoken from his ``dogmatic slumbers'' \citepalias[4:260]{kant1783} with regard to the principle of causality. However the legitimacy of the principle had been undermined only with respect to metaphysical cognition. Its use had not, in Kant's mind, been undermined within the domain of natural science. He writes, in 1783:

\begin{quote}
The question was not whether the concept of cause was right, useful, and even indispensable for our knowledge of nature, for this Hume had never doubted; but whether that concept could be thought by reason \emph{a priori}, and consequently whether it possessed an inner truth, independent of all experience \citepalias[4:258--259]{kant1783}.
\end{quote}

He continues: ``This was Hume's problem. It was a question concerning the \emph{origin} of the concept, not concerning its indispensability in use. Were the former decided, the conditions of its use and the sphere of its valid application would have been determined as a matter of course'' (ibid.).

Synthetic a priori cognition (a term Kant introduces in the first Critique) is that cognition in which two (or more) concepts are cognised, in advance of experience, as necessarily connected in some way to one another. `Ampliative' cognition of this kind is what metaphysics seeks. But by the time of the Critique, Kant is convinced that such cognition is impossible without a reference to the forms of our sensibility, space and time.\footnote{Cf. A5/B9.} Even the principle of causality has validity only with regard to appearances, which are always given in space and time. Thus, once we carry out what is required by the Dissertation's principle of reduction, metaphysical cognition\textemdash cognition of the intelligible world, i.e., of things as they are independently of \emph{all} conditions for the (possible or actual) experience of them\textemdash must be given up forever.

This does not undermine, however, the idea that the concept of cause is indispensable in some sense, nor does it imply that all synthetic a priori cognition is impossible. Thus Kant's project now becomes to explicate that sense and to show \emph{how} synthetic a priori cognition is possible and to investigate its limits. These limits are \emph{possible} experience, and since the \emph{form} of possible experience is given a priori, synthetic a priori knowledge is possible regarding it. A useful \emph{corollary} to the results of this investigation is a grounding for the theoretical sciences; i.e., an answer to the issue that Hume \emph{did not} intend to raise: that of the validity and scope of principles, such as that of causality, \emph{within} the theoretical sciences.\footnote{Kant writes, at B20: ``In solving the above problem we solve at the same time another one, concerning the possibility of the pure use of reason in establishing and carrying out all sciences that contain theoretical a priori cognition of objects''. Cf. also: \citetalias[4:280]{kant1783}.} Metaphysics, regarded as a system of synthetic a priori constitutive principles which \emph{transcend} experience, is declared impossible. Theoretical metaphysics in the traditional sense is instead transformed into a system of methodological principles for the investigation of nature.\footnote{At least part of the reason for the second edition emendations of the first Critique seems to have been to clarify this goal of delimiting the domain and conditions for the possibility of synthetic a priori cognition. See, for example, the Feder-Garve review \citep[pp. 53-58]{sassen2000}, and Kant's response \citepalias[4:372-380]{kant1783}.}

The `useful corollary' just mentioned---i.e. the grounding for the theoretical sciences that falls out of an investigation into the scope and limits of synthetic a priori cognition---is primarily found in the Transcendental Analytic section of the Critique, and especially in the Analytic of Principles. Recall from \S\ref{sec:negalt} that for Kant, possible experience comprises two distinct aspects: intuition, for which space and time are its pure forms, and \emph{concepts of the understanding}, which correspond to rules for synthesising the manifold of intuition. The concept `chessboard', for instance, corresponds to a rule whereby this particular bit of white, that particular bit of black, etc. can be associated in one representation. When we synthesise, i.e. combine, some particular manifold of intuition according to the particular rule for a concept, we say that the manifold has been subsumed under the concept and cognised in accordance with it (A68-69/B93-94). A \emph{pure} concept of the understanding is called a \emph{category} (A80/B106); it is one of a set of `meta-concepts' that all empirical concepts necessarily presuppose. Like the pure forms of intuition, the categories are a priori.

The application of a category to the manifold of intuition is governed by the \emph{schema} of that category (A137-147/B176-B187); i.e. the rule by which the manifold is determined in accordance with the category. For example, the pure schema of magnitude is number (A142-143/B182). In apprehending the manifold corresponding to an object, determinate intuitions---`instants'---of time are produced with each successive act of synthesis. Through these is produced a \emph{time series} from which one judges as to the extent of the object apprehended (A145/B184). Through an analysis of the use of the categories in accordance with their characteristic schema, we are entitled to make a number of synthetic a priori judgements regarding the objects of cognition in general. These are called synthetic a priori \emph{principles} (A159-235/B198-294) by Kant. Of particular importance for our discussion in the next and subsequent sections, wherein we will be considering the relevance of quantum theory for Kant's philosophy, is the principle of causality. Kant calls this a `dynamical principle' since it is a principle governing the connection of appearances in time. It tells us, according to Kant, that the changes undergone by an object of cognition are ordered uniquely and objectively according to a necessary rule (A191/B236). This is in contrast to the series of subjective perceptions of the object through which we apprehend it.

To illustrate: suppose I lean against a fence at the bank of a river, and watch a piece of wood as it is carried downstream by the current (cf. A191-193/B236-238). At time $t_1$, I watch as it comes into view from around the bend in the river some yards upstream. I then daydream for a while. Eventually, I notice (at $t_2$) that the log has travelled some distance from the place where I first spotted it. At $t_3$, I recall to myself the motion of the wood down the river that I half-consciously observed while daydreaming, after which I continue to watch as it disappears into the forest ($t_4$). Later that afternoon, I recall that what aroused me from my daydream was a sparrow alighting on the log (at $t_5$). If we list these representations in the order in which they are actually perceived, then this is a \emph{subjective ordering}: $$t_1, t_2, t_3, t_4, t_5.$$ I can also give them an \emph{objective ordering}, however, according to which the motion of the log must have actually proceeded in time: $$t_1,t_3,t_5,t_2,t_4.$$ To determine this objective ordering, I must discover a rule for the log's motion, for in general ``appearance, as contrasted with the [representations] of apprehension, can be [represented] as an object distinct from them only if it is subject to a rule that distinguishes it from any other apprehension and that makes necessary one kind of combination of the manifold.'' (B236/A191).\footnote{I have modified Pluhar's translation of \emph{Vorstellung} as `presentation' to the more standard `representation'. As \citet[p. 348]{allison2001} points out, Kant considered this term as equivalent to the Latin \emph{repraesentatio} (A320/B376).} The particular rule of succession for the change of state of the log is something that can only be discovered empirically, for instance by taking into account the position of the log in the river in my various perceptions of it, the direction therein in which I perceived the river to be flowing, etc. However, that objective cognition \emph{presupposes} that there is some such rule is what the principle of causality tells us. This is \emph{a priori}, according to Kant. One notable aspect of such a rule, which we will return to later, is that Kant seems to require that it be deterministic in the sense of making possible the perfect \emph{prediction} of an event from what has immediately preceded it. In Kant's words: ``what precedes an event as such must contain the condition for a rule whereby this event always and necessarily follows.'' (B238--239/A193).

The concept of something that is an object \emph{for us}---i.e. of something we have achieved objective cognition of---presupposes a determination of the manifold in accordance with the principle of causality and the other synthetic a priori principles, for Kant (A159/B198). On the other hand, and importantly, these principles are only regulative with respect to experience `as a whole'. By this is meant that from a \emph{methodological} point of view we are required to investigate nature in accordance with them (for otherwise the objective cognition of anything would be impossible). However it does not follow from this that the objects of our inquiries are already determined in accordance with these principles in advance of our investigations, or that we can know a priori that any particular endeavour to attain objective cognition will be successful (cf. A509/B537). A particularly striking example of the failure to obtain objective cognition in this sense is provided by quantum theory, which we will begin to discuss in the next section.

\section{}
\label{sec:quantmech}

To begin with, consider the following `classical' scenario.\footnote{The following example is adapted from \citet[p. 381]{bacciagaluppi2015}, as is my discussion of the analogous single-slit example involving quantum phenomena which follows. The quantum example is of course originally due to \citet[]{bohr1935}. In his paper, Bohr actually discusses (successively) multiple variations of the single-slit experiment. The one discussed here and in \citet[]{bacciagaluppi2015} is the particular variation discussed on pp. 698--699: ``In an arrangement suited for measurements of the momentum of the first diaphragm, it is further clear that even if we have measured this momentum before the passage of the particle through the slit, we are after this passage still left with a \emph{free choice} whether we wish to know the momentum of the particle or its initial position relative to the rest of the apparatus''.} A medium-sized object is launched towards a diaphragm into which an opening, or `slit', has been made that is large enough to allow the object to pass through, but small enough so that it invariably deflects the object to some degree as it does so. After passing through the slit the object eventually impacts upon a further screen. Assume that the diaphragm is movable (e.g. let it be attached to the rest of the apparatus by springs), so that when the object collides with the edges of the slit as it passes through the diaphragm, the latter recoils slightly. We would like to describe the state of the object immediately after its interaction with the diaphragm. Assume that the (centre-of-mass) positions and momenta of both the object and the diaphragm just prior to the interaction are known, but that the precise shape and size of the launched object are unknown, so that we cannot calculate in advance what the respective positions and momenta of the object and diaphragm will be after the collision. Once they do collide, however, we can then measure the momentum of the diaphragm by observing the amount by which it recoils, which will allow us to (via the law of conservation of momentum) determine what the momentum of the launched object is immediately after the collision. And since the common centre of mass of the combined system remains at rest, by also measuring the position of the slit we are able to determine the object's position.

In classical physics, which is adequate to describe the motion of objects such as the one in the imagined scenario, a precise determination of momentum and position is sufficient to completely characterise the state of a particle (or collection of particles) \emph{vis \'a vis} its variable parameters at any one time. And from the precise characterisation of the state of a particle at any one time, one can then precisely infer the state of the particle at all other times, assuming the system is isolated or closed \citep[\S 2.1]{hughes1989}.\footnote{Note, however, Hughes's caveat in \S 2.4.} Thus through precisely determining the state of the object immediately after its impact with the diaphragm, we are able to predict the object's subsequent trajectory through its state space with certainty.

This is in spite of the fact that the diaphragm---our measuring instrument---disturbs the motion of the object as the latter impacts upon it. For in classical theory this is in principle unproblematic. In every case of classical measurement, either the interaction of the measuring instrument with the object of inquiry can be made negligible for the purposes of the analysis, or (as above) we can use physical theory to abstract away from this interaction in such a way as to allow us to precisely determine the object's various state parameters. Indeed for precisely this reason, after we have measured the momentum of the \emph{diaphragm}---which will in general disturb \emph{it}---we can subsequently measure the diaphragm's position, and then use physical theory to abstract away from the interaction involved in our previous momentum measurement to determine the position of the object which has just passed through. In this way we are able to simultaneously ascribe \emph{both} position and momentum parameters---i.e. a complete state description---to the launched object \citep[]{bai2017}.

In the words of the previous section, in every case classical theory allows us to transition from the subjective conditions (represented by the diaphragm in the above example) under which we perceive an object, to an objective description of that object wherein these subjective conditions no longer explicitly appear. They nevertheless remain implicit in the sense that our description of the object presupposes that it has been determined by us to be such as we describe it (either directly or perhaps only indirectly) through a process of measurement in space and time. This process of determination is characteristic of \emph{all} objective cognition, as we have seen, for Kant.

Things are more interesting when we come to the case of quantum phenomena. Let the `launched object', i.e. system of interest, now be quantum---a photon, for example---which passes through the slit in a diaphragm on its way to an eventual impact with a photographic plate. Similarly as in our previous example, we would like to objectively describe its state immediately after it has collided with the slit. In this case, however, quantum mechanics' well-known `uncertainty'---or as Bohr preferred to call it: `indeterminacy'---relation for position and momentum precludes us from ascribing determinate values to these quantities simultaneously. For according to this relation, as the indeterminacy in the position of an object approaches zero, the indeterminacy in its momentum approaches infinity, and vice versa. Expressed in terms of our example, this means that if we choose to precisely measure the momentum of the diaphragm (so as to ascribe a determinate momentum to the object that has just passed though it), then the consequent disturbance of the diaphragm will be such as to make impossible a further precise determination of the object's position. That is, the diaphragm's displacement consequent upon the momentum measurement of it will have been, in contrast to the situation in classical theory, ``uncontrollable'' \citep[p. 698]{bohr1935}---or at any rate not controllable enough---and we will be unable to account for this displacement and abstract away from it as we did in the classical case for the purposes of a subsequent position determination of the object. The situation will be similar if we instead choose to set up the experiment so as to make determinate the diaphragm's position (and thus the position of the system of interest); through this choice we will have precluded ourselves from precisely determining the system's momentum.\footnote{\label{fn:micro1}Note that if we choose not to measure the diaphragm at all, then the joint state of the diaphragm and photon will be describable as an entangled quantum superposition. \citet[\S 10]{hermann1935a} makes this point (implicitly) in the context of a different thought-experiment (the $\gamma$-ray microscope experiment). The role played by the diaphragm in our example is in that context played by a photon which is collided with an electron (the system of interest) in order to determine the latter's state. In place of a pointer connected to a diaphragm, we have in that context a photographic plate which can be set up in various ways (or not at all) in order to measure the state of the photon. See also fn. \ref{fn:micro2} below.}

\section{}
\label{sec:conceptind}

Bohr expresses the significance of the limitation imposed by the indeterminacy relations as follows:

\begin{quote}
Indeed we have in each experimental arrangement suited for the study of proper quantum phenomena not merely to do with an ignorance of the value of certain physical quantities, but with the impossibility of defining these quantities in an unambiguous way \citep[p. 699]{bohr1935}.\footnote{An earlier statement expressing an essentially identical viewpoint can be found in \citet[p. 580]{bohr1928} and is discussed in \citet[pp. 311--312]{cuffaro2010}.}
\end{quote}

That is, the significance of the indeterminacy relations is, for Bohr at any rate, not epistemological in the sense that one presupposes the quantum object in the above example to be perfectly determinate in itself with respect to all of its ascribable physical parameters, but yet not completely knowable by us.\footnote{This is, incidentally, the viewpoint normally associated with (the young) Heisenberg, although see the discussion in \citet[pp. 92-94]{frappier2017} for a contrary view.} What is being expressed here, rather, is that their significance is epistemological in a different sense; a better word would be \emph{conceptual}.

From the point of view of \S\ref{sec:synthap}, we can understand this as follows.\footnote{I consider Bohr's views on quantum mechanics to have substantial elements in common with Kant's philosophy, and I will be pointing out some of these elements as we proceed. However my goal in this paper is not specifically to present an argument for this interpretation of Bohr, as I take that to have already been established by the work of numerous others (some of whom have been referred to above). Indeed, even those who have identified pragmatism as the more dominant influence on Bohr's thought now acknowledge that the influence on him of Kantianism was nevertheless substantial \citep[e.g.][]{camilleri2017, faye2017a, folse2017}. In any case, as I alluded to in the introduction to this paper, and in agreement with \citet[]{folse2017}, I do not see pragmatism as incompatible with Kantianism when the latter is construed broadly in the sense of a research program \citep[cf.][]{bitbol2017}.} Consider the result of some experiment, say the mark on a photographic plate, or the particular situation of a pointer measuring the momentum of the diaphragm in the above example. The pointer and the mark themselves---considered simply as a mark or as a pointer, respectively---are describable as classical objects in the sense that they can each when observed be described in the ordinary way as having definite spatiotemporal coordinates, and as causally interacting in a definite way with their surroundings. However our aim (which as we will see cannot be satisfied in quite the way one would like in the quantum context) is to go beyond the particular mark and the particular pointer reading and describe these as having arisen through the interaction of our experimental apparatus with some independently existing object. Our goal, in other words, is to `get at' this object as it exists independently of the `subjective conditions' associated with the particular experiments we subject it to. And the way we do this---or anyway the way we attempt to do it---is by compensating mathematically for the interaction between the apparatus and object in our description of the latter.

For a Kantian, it is of course never possible to \emph{completely} eliminate in this way the subjective conditions pertaining to our observation of an object from our objective description of it, since any description of an object that is cognisable for us must be determinable with respect to space and time. This is perfectly acceptable, however, since our standard for objective cognition is anthropocentric rather than theocentric. And since space and time are the pure forms attaching to any possible experience, we can in principle achieve a kind of objectivity through the attainment of a description of the object which will be valid for all discursive cognisers such as ourselves.

Now in order to describe something objectively in this sense---i.e. as an object of possible experience that exists independently of us in the sense that its description does not depend on the particular way that we \emph{actually} come to know it---it must be determinable in space and time in accordance with the synthetic a priori principles. Earlier I mentioned causality as an example of a `dynamical' principle of this kind. In addition to the dynamical principles there are also what Kant calls \emph{mathematical} principles (B198--B294).\footnote{The mathematical principles are the \emph{Axioms of Intuition} and \emph{Anticipations of Perception}; the dynamical principles are the \emph{Analogies of experience} and the \emph{Postulates of empirical thought as such}.} According to the latter, anything that appears to us must be apprehended as having, determinately, both an extensive (length, breadth, etc.) and an intensive magnitude (i.e. a degree). The dynamical principles, in contrast, are not principles for the apprehension but for the connection of appearances in time. They state, first, that all change presupposes something permanent; second, that all change must occur according to the law of cause and effect; third, that all substances that are perceived as simultaneous are in mutual interaction.\footnote{Here, I only consider the \emph{Analogies}, as the \emph{Postulates} are not directly relevant for our discussion.}

An objective description in the above sense is one that is determined in accordance with \emph{both} sets of principles. Together, they assert that the determination of any appearance as an object of possible experience must be such that at a determinate instant in time, it has a determinate extent (constrained by the mathematical principles) and hence a determinate position in space, and that there is a law (subject to the dynamical principles) by which it dynamically interacts with its surroundings in and through time. In the context of our example of the slit, one can interpret this as signifying that any description of a quantum object that purports to pick out an object of possible experience for us must be such as to ascribe to that object both a determinate position and a determinate momentum parameter. But according to the indeterminacy relations, it is impossible in principle to describe the particle's momentum with any degree of precision without a corresponding loss of precision with regards to its spatial coordinates. The upshot of all of this is that we cannot complete our description of the object according to the Kantian criteria for objects of possible experience. And yet these are \emph{necessary} criteria, for Kant, in the sense that, as we have seen, objective knowledge is impossible for us without them. We may get around this in some sense by ascribing only indeterminate values of position, momentum, etc. to the `object' of our investigations, but the resultant `unsharply defined' description \citep[cf.][p. 582]{bohr1928} can as a result never be an object \emph{for us}---i.e. it can never be an object which we can have possible experience of and thus never be real for us in that sense. We can consider it merely as a noumenon, or abstract object \citep[see][p. 313]{cuffaro2010}.

For a Kantian, the situation thus seems hopeless. An objective description just is one in which we determine something according to both spatiotemporal and dynamical criteria---we simply have no other choice. And yet these criteria cannot fulfil their intended function in the quantum domain, for a determination of one necessarily excludes a determination of the other in the sense of the indeterminacy relations. Further, there is a different (though related) sense in which they mutually exclude one another as well, which stems from the so-called `wave-particle duality' of quantum phenomena. As Bohr points out \citep[p. 581]{bohr1928}, in the equation expressing this duality: $E\tau = I\lambda = h$, Planck's constant ($h$) relates quantities that are incompatible from a classical point of view. That is, in the first relation, $E$ (energy) is associated with the concept of a particle given with definite spatiotemporal coordinates, while $\tau$ (the period of vibration) is associated with a wave-train ``of unlimited extent'', not conceptualisable with respect to definite space-time coordinates. Likewise, respectively, for $I$ (momentum) and $\lambda$ (wavelength). Bohr's point is that it is inconsistent from a classical point of view to describe an object as being, in accordance with the above relations, both given at some definite spatiotemporal location and of unlimited extent in space and time. However a violation of the indeterminacy relations in the description of a quantum object would imply \citep[p. 311]{cuffaro2010} a precise determination of that object with respect to both of the above parameters, which would entail the simultaneous applicability of these mutually exclusive conceptualisations of the object, which cannot be (i.e. on pain of contradiction). Thus not only do the classical---or Kantian for our intents and purposes---criteria mutually exclude one another \emph{vis \'a vis} their determinability. Even if we could determinately ascribe both spatiotemporal and dynamical attributes to the object, the resulting object would be self-contradictory.

Ironically it is the uncertainty relations which save us, at least to some extent. They guarantee that we can nevertheless achieve a unified---albeit abstract---description of quantum phenomena by `patching together' the mutually exclusive dynamical and spatiotemporal descriptions that result from our various experiments. As Bohr puts it:

\begin{quote}
The apparently incompatible sorts of information about the behaviour of the object under examination which we get by different experimental arrangements can clearly not be brought into connection with each other in the usual way, but may, as equally essential for an exhaustive account of all experience, be regarded as ``complementary'' to each other \citep[p. 291]{bohr1937}.
\end{quote}

The uncertainty relations guarantee that a dynamical description can never contradict a spatiotemporal description\textemdash that the two can be used in a complementary way in our description of an abstract quantum object\textemdash for any experiment intended to \emph{determinately} establish the object's spatiotemporal coordinates \emph{can tell us nothing} about its dynamical parameters, and vice versa.

\begin{quote}
the proper r\^ole of the indeterminacy relations consists in assuring quantitatively the logical compatibility of apparently contradictory laws which appear when we use two different experimental arrangements, of which only one permits an unambiguous use of the concept of position, while only the other permits the application of the concept of momentum ... \citep[p. 293]{bohr1937}.
\end{quote}

We are not licensed, however, to take the next step and ascribe physical reality to this `patched together' object of our descriptions, for the object is not real but abstract, and its classical spatiotemporal and dynamical attributes are idealisations.

\section{}
\label{sec:schlick}

For one seeking to defend the Kantian philosophical framework, which includes the Kantian synthetic a priori principles---and in particular the principle of causality---this is good news, at least in one sense. For despite the fact that the indeterminacy relations imply that in general it is impossible to predict with certainty the future behaviour of a quantum object from a complete characterisation of its present state, i.e that the latter in general does not ``contain the condition for a rule whereby [a subsequent] event always and necessarily follows'' (B238--239/A193--194), Kant's synthetic a priori principles are not thereby invalidated. For quantum state descriptions are merely abstract objects from a Kantian point of view and thus are not objects of possible experience for us. It is only the concept of the latter which presupposes that it be determinable in accordance with the principle of causality and other synthetic a priori principles. And it is only for such objects of possible experience that the perfect prediction of its subsequent states from a determinate description of the object is implied.

There is, nevertheless, a potential problem, if not for Kant's framework as a whole then at least for the principle of causality. For as we saw, the principle of causality is, in addition to being presupposed by the concept of objective cognition, also (arguably for precisely this reason) a regulative principle for the purposes of the investigation of nature generally speaking. But if objective cognition in the Kantian sense is typically excluded in quantum mechanics, and if moreover the doctrine of complementarity allows us to continue to do physics in spite of this, then it seems that we can do away with causality as a regulative principle even if, strictly speaking, it is not contradicted by quantum mechanics. The logical empiricist Moritz Schlick expressed this anti-Kantian objection in 1931 as follows:

\begin{quote}
The principle of causality does not directly express a fact to us, say, about the regularity of the world, but it constitutes an imperative, a precept to seek regularity, to describe events by laws. Such a direction is not true or false but is good or bad, useful or useless. And what quantum physics teaches us is just this: that the principle is \emph{bad}, useless, impracticable within the limits precisely laid down by the principle of indeterminacy. Within those limits it is impossible to seek for causes. Quantum mechanics actually teaches us this, and thus gives a guiding thread to the activity that is called investigation of nature, an opposing rule against the causal principle. \citeyearpar[p. 285]{schlick1962}.
\end{quote}

This objection strikes deep---indeed to the very heart of Kant's philosophical project---which as we saw was, at least as early as 1755, the methodological one of providing a principled means for acquiring objective cognition---of providing the ``materials for constructing a building'' as he put it in 1763 (2:66). For if the Kantian principles are inapplicable within a particular domain of physics, it can be but small comfort to know that they are (for that reason) still strictly speaking valid in general. Kant's epistemological grounding, this objection asserts, is irrelevant within this specific domain and of no use to us.

By way of response, one might note that Schlick's construal of the situation regarding the principle of causality in quantum mechanics only applies to the quantum state description itself, i.e. it shows us that our `patched together' description of a quantum object is not the description of an object of possible experience and thus not subject to the principle of causality in that sense (as we saw in the previous section). But while it is true that one cannot simultaneously attach determinate spatiotemporal and dynamical attributes to this description, it remains the case on the viewpoint expressed here that we anyhow require the Kantian categories and principles in order to `patch together' this very description. That is, they are necessary for the very interpretation of the results of the measurements involving mutually exclusive experimental arrangements from which we construct our description of the quantum object. As Bohr puts it:

\begin{quote}
Here, it must above all be recognized that, however far quantum effects transcend the scope of classical physical analysis, the account of the experimental arrangement and the record of the observations must always be expressed in common language supplemented with the terminology of classical physics. \citep[p. 313]{bohr1948}.

\vspace{12pt}

The main point here is the distinction between the \emph{objects} under investigation and the \emph{measuring instruments} which serve to define, in classical terms, the conditions under which the phenomena appear. \citep[pp. 221-222]{bohr1949}.
\end{quote}

We require the classical concepts, not only to observe, but also to communicate experimental results:

\begin{quote}
... the requirement of communicability of the circumstances and results of experiments implies that we can speak of well defined experiences only within the framework of ordinary concepts \citep[p. 293]{bohr1937}.
\end{quote}

In this sense the `classical concepts' do in fact continue to function at least as methodological \emph{tools} for the investigation of nature. The epistemological lesson driven home by quantum mechanics, one might say, is that epistemic primacy does not imply ontological primacy; viz., the necessary principles for cognising something as an object of possible experience cannot be construed as necessary conditions for the possibility of objects in themselves. But this is something we (as Kantians) should already know, for it is what is asserted under an epistemic interpretation of Kant's transcendental idealism. That is, it \emph{does not follow} from the fact that we necessarily apprehend the world in this way that it must conform to our way of apprehending it. In the context of Kant's synthetic a priori principles this means that it does not follow from the fact that we are required to investigate nature in accordance with these principles that the objects of our inquiries are already determined in accordance with them in advance of our investigations, nor that we can know a priori that any particular endeavour to attain objective cognition will be successful (cf. A509/B537). These principles are constitutive for objects, but only regulative for nature as a whole.

But how, one might ask, is this `patching together' of the information regarding the object which we glean from our various experiments in accordance with the classical concepts actually accomplished? The indeterminacy relations show us only that (as emphasised by Bohr) this patching together cannot lead to contradictions, but it remains to be clarified just how the classical concepts can be used in this way. This is, in part, the contribution of Grete Hermann. And as we will see in the next section, in addition to this she lays down an alternative---but still broadly Kantian---understanding of the lesson provided to philosophy by the discoveries of quantum theory, which can be seen as stemming from a Friesian-Nelsonian conception of the fundamental nature of cognition.

\section{}
\label{sec:hermann}

Hermann's reflections on quantum mechanics and its significance for the `critical philosophy'---the philosophical tradition begun by Kant and continued by Fries as well as by her own teacher, Leonard Nelson---are for the most part contained in a long essay that she published in 1935 \citep{hermann1935a}.\footnote{For more on Grete Hermann's life and intellectual background see, for instance, \citet[]{hansenSchaberg2017}, \citet[]{leal2017}, and \citet[]{paparo2017}.} She begins this essay with the admission that the emergence of quantum theory has ``shaken'' (p. 239) this tradition and in particular has shaken the idea that a priori principles discoverable through critical philosophical analysis lie at the foundations of natural science. Ultimately she will argue that the challenge posed by quantum theory to the critical philosophy can be met. However it cannot be met, she argues, simply by recapitulating the critical deduction that had previously been made for these principles.\footnote{As we will see in more detail later, `deduction' is being used here in a specific technical sense that is relevant to Friesian critical philosophy. Throughout her essay, Hermann generally presupposes her readers to be familiar with Friesian doctrine including its characteristic jargon (her essay was originally published in the Friesian journal \emph{Abhandlugen der Fries'schen Schule}).} She writes:

\begin{quote}
For, even if the physical development of the theory is not sufficient to put the foundations of the thus achieved knowledge of nature into the sharp light of awareness, still the scientific progress that has been obtained in these theories precisely through the willingness to abandon or revise old familiar concepts provides the guarantee that new and fruitful points of view have been introduced here into research. Only their philosophical interpretation and elaboration will produce clarity concerning both the philosophical arguments for the a-prioricity of natural-philosophical principles and the objections to them arising from the side of physics. \citep[p. 240]{hermann1935a}.
\end{quote}

In looking to quantum theory to illuminate the critical philosophy in this way, Hermann's attitude differs starkly from that of her teacher Nelson. Nelson, as we will see later, was far from denying the general idea that philosophical insight is attainable through an analysis of scientific knowledge. However Nelson held that the physics of the early twentieth century (not just quantum theory) was still too immature and full of contradictions to provide the kind of insight required to illuminate the critical philosophy. He rather recommended that---for the time being---critical philosophers adopt ``the self-denying stance of Conventionalism'' \citep[p. 253]{nelson1971} toward physics and consider it to be ``of purely heuristic significance'' (ibid.) for the purposes of critical philosophy.

But while Hermann maintained that real insight into the a priori philosophical principles of natural science could be gained through a critical philosophical analysis of the challenge posed to them by quantum theory, she noted that this could not take the form suggested by physicists such as Born, i.e. of attempting to disprove quantum theory by empirical means so as to restore these principles to their former unexceptioned status \citep[]{born1929}. This is because the question raised by quantum theory is whether such principles are a priori precisely in the sense of being amenable to a purely philosophical, albeit in some sense empirically informed, analysis \citep[pp. 240--241]{hermann1935a}.

With respect to the principle of causality in particular, Hermann also argues (in agreement with Schlick\footnote{Cf. \citet[\S 6]{schlick1961}.}) that it would be pointless, in the face of the challenge posed by the indeterminacy relations, to abandon the criterion through which causal connections in nature can be known; i.e. the criterion of prediction. She writes:

\begin{quote}
One who wished to brush this off with the excuse that, while the \emph{knowledge} of the causes determining the processes is limited, the \emph{existence} of such causes is not put in doubt, removes the law of causality from the realm of the principles governing natural knowledge into that of mysticism. Where it is impossible in principle to decide what falls under a given concept in nature, the statement \emph{that} anything falls under it also loses its meaning. \citep[p. 242]{hermann1935a}.
\end{quote}

Hermann nevertheless cautions against the use of the criterion of prediction in the ``positivistically distorted form'' (ibid.) employed by certain philosophers of her day. Presumably she has in mind those, such as Schlick, who simply identify causality with prediction. For Hermann, apparently, being able to satisfy the criterion of (perfect) prediction is necessary, in some sense, in order to have a meaningful notion of cause. But the concept of causal connection is for Hermann nevertheless quite distinct from the concept of prediction. The latter functions as the criterion by which we identify causes, but does not in itself express the concept of causal connection.

Now our application of the criterion of prediction is limited within quantum mechanics by the indeterminacy relations. But for Hermann the significance of this limitation is not that quantum objects always possess joint determinate values of, say, position and momentum that we must necessarily remain ignorant of to some extent. Rather, for Hermann their significance consists in the fact that the simultaneous subsumption of our description of a quantum object under the wave and particle conceptualisations of phenomena is only possible via limited applications of these two pictures to the phenomena \citep[p. 246]{hermann1935a}. Hermann's view is in this way similar to Bohr's. However, as we will see shortly, Hermann's understanding of the wider significance of this circumstance marks a subtle but important difference with Bohr that is relevant to our earlier discussion of transcendental idealism.

In any case, precisely because a quantum object cannot, according to these relations, be simultaneously described as having a determinate position and momentum, it follows that it cannot be the case that its future motion is in principle determined by (and hence perfectly predictable from) these state parameters as it would be for a classical object. But this raises the question of whether there may be further (testable) parameters, not described by quantum theory, which in fact precisely determine the quantum object's motion. This question, Hermann maintains, is a meaningful and legitimate one to which a positive answer is conceivable despite the various so-called proofs (which she argues are unsound) purporting to show the impossibility of such parameters in general \citeyearpar[p. 254]{hermann1935a}.\footnote{Notable is her analysis \citeyearpar[pp. 251-253]{hermann1935a} of von Neumann's impossibility proof which anticipates the famous objections of Bell \citep[see][]{seevinck2017}. Note that the argument presented in her 1935 paper is mostly a reproduction of the arguably clearer version from her earlier unpublished paper \citep[]{hermann1933}.} Indeed, for Hermann ``there can be only one sufficient reason for abandoning as fundamentally useless the further search for the causes of an observed process: \emph{that one already knows these causes}.'' \citeyearpar[p. 254]{hermann1935a}.

Nevertheless (and perhaps surprisingly) Hermann maintains that quantum mechanics gives us no reason to continue the search for additional causes of the processes it describes, since quantum mechanics itself already provides the resources with which to identify them. In particular, Hermann notes that if one interprets the particular situation of a measurement pointer as a statement about the current state of a quantum system of interest, this interpretation presupposes a (classical) theory of the interaction that has taken place between the quantum system and the measuring apparatus through which the particular situation of the pointer has arisen.\footnote{Compare this with the statements attributed to Einstein in \citet[pp. 63-64]{heisenberg1971}.} For instance, in the above example of the diaphragm and slit one appeals to the conservation law for momentum in order to interpret the reading of the pointer after one has measured the diaphragm as asserting a fact about the quantum object which has just passed through it. But in making such an inference, Hermann argues, the reading of the pointer is thereby ``explained as the necessary effect that the system to be measured has imposed on the instrument in the process of measurement'' \citep[p. 255]{hermann1935a}. In other words we posit a fact about the quantum system of interest---that it has a particular momentum---as an explanation which, in light of the classical conservation theorems, accounts for the fact that the pointer reading has a particular value. Moreover such a posit is testable, for quantum theory predicts that, given this posit regarding the system of interest's momentum, a subsequent measurement of that parameter will (experimental error aside) yield the same value again with certainty.\footnote{This is just the statement that quantum measurements are repeatable.}

Because of this there is no need, Hermann claims, to seek for the physical features overlooked by quantum mechanics which would make possible a causal explanation of the measurement pointer's particular reading. For a (classical) causal explanation for this particular measurement result is already provided by quantum mechanics in just the way mentioned. Likewise, the classical theory of interaction appealed to in this case also provides a classical causal explanation for the particular value we ascribe to the system of interest itself subsequent to its interaction with the diaphragm. Nothing about this analysis changes essentially, moreover, if it is a microscopic object that plays the role for us of a measuring instrument \citep[p. 255]{hermann1935a}, for example when we use a microscope to collide a photon with an electron in order to measure the state of the latter.\footnote{\label{fn:micro2} In \S 10 of her exposition, Hermann explores the microscope example alluded to here at length. However Hermann's essential conceptual point, at least for our purposes, is made in the section we have been focusing on: \S 9. Besides being a concrete illustration of the discussion of \S 9, \S 10 additionally contains further contributions to quantum foundations. It contains, for example, one of the earliest discussions of quantum entanglement in either the physical or philosophical literature. A detailed analysis of this and of the particulars of Hermann's microscope example is inessential for our own discussion, however, thus we will not consider them here, and will for the most part continue our discussion in the context of Bohr's example of the diaphragm and slit. For more on the particulars of Hermann's microscope example, see \citet[]{filk2017, frappier2017}. A short summary of the relation between the two examples is given above in fn. \ref{fn:micro1}. For a more detailed discussion and comparison of the two examples, see \citet[]{bacciagaluppi2017a}.}

Returning to the example of the diaphragm and slit: Recall that due to the uncontrollable displacement of the diaphragm consequent upon a measurement of its momentum, we are (as we have already seen) unable to form a posit regarding the system of interest's \emph{position} immediately after having interacted with the slit. This results in a kind of `one-sidedness' in our characterisation of the system of interest. As Hermann puts it:

\begin{quote}
the quantum mechanical description by which, on the basis of some observation, a physicist determines his system, does not characterise this system completely and absolutely, but (so to speak) reveals only one aspect of it---precisely the aspect that presents itself to the researcher on the basis of the observation made here. \citeyearpar[p. 256]{hermann1935a}.
\end{quote}

To put it a different way: For any experiment performed on a quantum system of this kind one can, in the way described above, appeal to a classical law by which the reading of our measurement apparatus emerges as the necessary consequence of a posited feature of the quantum system measured. In that sense we classically explain the reading of the pointer via this posit and the relevant classical conservation law. However at least from a classical point of view such a description of the quantum system of interest is not a fully objective one, in the sense that it leaves out information which we require for a classically complete characterisation of the system's state. But as Soler and others have correctly pointed out,\footnote{See, especially, p. 65 of \citet[]{soler2017}. \citet[p. 257]{banks2017} makes a similar point, and so does \citet[p. 140]{bacciagaluppi2017a}, although in the latter case \citeauthor[]{bacciagaluppi2017a}'s focus is on this difference between Hermann's view and Bohr's rather than Kant's.} this presents a problem for one seeking to defend a strictly Kantian conception of causality. Hermann seems to depart substantially from Kant's conception, in fact. For recall that for Kant a causal process is one in which the sequence of appearances of an object can be connected together in time (see \S \ref{sec:synthap} above). But the appearance of an object is an appearance which can be given an objective description. And an objective description is one in which \emph{both} dynamical and spatiotemporal parameters can be ascribed to the object. Thus the causal explanations that Hermann claims can be reconstructed from our various observations of phenomena do not seem to be causal in a Kantian sense. One may doubt, therefore, whether Hermann's analysis of the situation in quantum mechanics, even if one deems it to be a success otherwise, can be understood as a vindication of Kant's critical philosophy.

To be fair to Hermann, however, she nowhere claims to adhere completely strictly to the critical philosophy as it was expounded by Kant himself. Indeed, as we will see further below, she is in fact critical of Kant in her essay. We will also see how the conception of causality that results from Hermann's analysis of quantum theory can actually be seen as a natural consequence of applying the critical method as it was elaborated by Fries and Nelson---as a natural extension of Friesian neo-Kantianism, if you will.

\section{}
\label{sec:fries-nelson}

Fries, like Kant, saw the goal of critical philosophy to be primarily one of explicating the a priori in cognition, where the norm or standard for cognition is understood (to use Allison's terminology) not in a theocentric but in an anthropocentric sense. Recall that for Kant, the nature of our cognition is discursive. By this it is meant that cognition for us is composed of both a sensible and a conceptual aspect---the faculties of sensibility and understanding, respectively. Kant also refers to the faculty of \emph{reason}, and indeed dedicates the lion's share of the Critique (the Transcendental Dialectic) to this faculty. But by reason he means, according to Fries, merely dialectical reason, i.e. the ability to make inferences. For Fries dialectical reason is not essentially different from understanding; for understanding is the faculty of judgement, and dialectical reason is merely a kind of judgement. For Fries, understanding and reason in this sense should rather be thought of as together comprising a single faculty, namely the faculty of mediative or reflective reasoning. But in addition to this faculty and the faculty of sensibility, Fries holds that there must be a third autonomous faculty involved in our cognition. This is pure reason properly speaking, which operates spontaneously and involuntarily and is, for Fries, the proper source of the a priori principles of our cognition \citep[pp. 227--228]{leary1982}.

These a priori principles, for Fries, are immediate, though they are not sensations. They are cognitive, and yet they are not a species of reflective cognition. They thus constitute a kind of \emph{immediate cognition}, which when applied in particular instances manifests itself in what Fries calls a `feeling-for-truth' \citep[p. 178--179]{nelson1971}. However because the cognition represented by these a priori forms is not reflective cognition, it cannot be demonstrated via deductive proofs in the way that Kant had, according to Fries, aimed to do in the Critique. Rather, these forms of cognition may only be `deduced'---i.e. discovered and described---through an analysis of the subjective contents of our cognition. In this sense the method of the critical philosophy is, according to Fries, \emph{empirical}. Through a philosophical analysis of the `data' that is our subjective consciousness, we `deduce' the a priori forms of knowledge that are implicitly relied upon therein \citep[pp. 228--229]{leary1982}.\footnote{For a detailed account of Fries's method of `deduction' see \citealt[pp. 164--196]{nelson1971}. We will be considering part of Nelson's exposition in more detail below.} In this sense Fries's conception of critical philosophy can be thought of as, as Nelson puts it, a ``combination of Kant's notion of the Critique of Reason with Plato's idea that philosophical knowledge is fundamentally obscure'' \citep[p. 160]{nelson1971}.

The reader will at this point be forgiven for thinking that Fries's critical philosophy is itself obscure, however. The idea of a `feeling-for-truth' is unclear and seems to amount to a bald `appeal to intuition' characteristic of pre-Kantian philosophy or even of contemporary analytic metaphysics. In this sense it does not seem to have much in common with the science that is Kantian philosophy. More importantly the idea of an immediate cognition which, despite its immediacy, is only known obscurely seems incoherent \citep[cf.][p. 78]{beiser2014}. And it seems problematic to hold that knowledge of the a priori can be grounded empirically. It is thus quite fortunate that one of Nelson's most important contributions to Friesian critical philosophy was to clarify these and other aspects of Fries's views.

The distinction between the a priori forms of knowledge (what, in Nelson's terminology, is called the `system of metaphysics', by which he does \emph{not} mean metaphysical cognition in the sense of Kant's \emph{Inaugural Dissertation}) and the Friesian critical philosophy, Nelson explains, can be understood in terms of what we would today call the distinction between object-language and meta-language \citep[pp. 184--190]{nelson1971}. Nelson does not use exactly this terminology, though he comes close, and in any case expresses the essential ideas:

\begin{quote}
The peculiarity of the Critique of Reason, as practised by Fries, lies in this: the knowledge contained in the system of metaphysics forms the object of the knowledge contained in the Critique of Reason. \citep[p. 184]{nelson1971}.
\end{quote}

This is interesting and noteworthy, since the lecture notes from which these words are quoted (published posthumously in German in 1962 and then in English in 1970-71) date from the years 1919-1926---significantly earlier than Tarski's and Carnap's introduction and formalisation of these concepts in 1933 and 1934, respectively.\nocite{tarski1956, carnap1937} In any case Nelson illustrates the distinction via a discussion of the axiomatic method in mathematics. In the system of mathematical knowledge that we call geometry, Nelson explains, one often needs to answer questions about the various statements in the system and their relations to one another. To do so one uses what we would now call a meta-language, within which one is able to formulate assertions regarding what we would now call the object-language (in this case the system of geometrical statements). In this way one can express propositions about the object-language in the meta-language such as: `$\phi$ is unprovable within the system $\mathfrak{G}$' (where $\phi$ is a statement in the object language $\mathfrak{G}$), `The minimal set of axioms for $\mathfrak{G}$ is $\{\phi,\psi,\chi\}$', and so on \citep[p. 185]{nelson1971}.

To illustrate how the knowledge contained in the meta-language can be acquired empirically, suppose that one has not been given a complete specification of the system of interest (the object-language) at the beginning of one's investigation into that system.\footnote{This illustration of how one may learn about an axiomatic system `via observation' is not actually a part of Nelson's discussion, but it is fully in the spirit of that discussion.} Suppose instead that one is successively presented a series of theorems in the object language. In such a case it is evident that as each new theorem of the object language is exhibited, one's understanding of the object language---which one encodes within one's meta-language---will grow; we learn empirically, so to speak, about the characteristics of our object-language (e.g. its axioms).

This relationship between `meta-language' and `object-language' is entirely analogous, Nelson maintains \citeyearpar[p. 185]{nelson1971}, with the relationship between Fries's critical philosophy and the `system of metaphysics' (which in the present context is, as we noted above, just the system comprising the a priori principles of cognition). But in place of mathematical statements, our object-language will in this case contain statements such as `$B$: Every change has a cause'. Within the meta-language, on the other hand, analogously to the mathematical meta-language which contains statements such as `$\phi'$: $\phi$ is unprovable', the critical philosophy will contain a statement asserting: `$B'$: $B$ recapitulates some item of immediate knowledge', where $B'$ is `deduced' in the Friesian sense by way of the empirical knowledge we have acquired thus far.

Fries's tripartite conception of cognition amounts, as does the orthodox Kantian bipartite discursive conception, to a kind of ontological posit. However in both cases these are posits regarding the concept of cognition. They do not aim to make substantive assertions about the properties of objects in the mind-independent world (cf. \citealt[][p. 166]{nelson1971}, \citealt[\S 12]{beiser2014}). As for the actual a priori principles of cognition which are discovered through the Friesian critical process of `deduction', these in the end turn out to be identical to those arrived at by Kant. In this sense one may say that Fries's contribution to critical philosophy pertains not to the elaboration of the end result of critical investigation but rather to the method that is to be followed so as to arrive at that result. Moreover since this critical method is, for Fries, an empirical one, it seems that it cannot be ruled out that further investigation will uncover new a priori forms or even modifications in our understanding of the a priori forms we have already uncovered. \emph{The critical method may in principle yield surprises.} And this, Grete Hermann maintains, is precisely what occurs when we re-examine the critical philosophy in the light of the discoveries of quantum mechanics.

\section{}
\label{sec:herm-transcideal}

Hermann begins her discussion of the deeper consequences for the critical philosophy of the discoveries of quantum mechanics in \S 16 of her 1935 essay. She begins with the observation that the nature of the Kantian system of natural philosophical principles (i.e. his synthetic a priori principles) is intimately bound up with his doctrine of transcendental idealism, which she characterises as follows: ``according to [transcendental idealism, the synthetic a priori principles] cannot provide adequate knowledge of reality `in itself' but only a limited knowledge of nature that stops at the conceiving of `phenomena'.'' \citep[p. 271]{hermann1935a}. Hermann then considers two Kantian arguments for transcendental idealism. The first begins from what we referred to above as the subjectivity thesis: that space and time are the formal and necessary subjective conditions for appearances. She takes Kant's argument for transcendental idealism from subjectivity to be unsuccessful, interestingly, for the same reason that most ontological interpreters of transcendental idealism do, namely because it does not follow from the fact that space and time are the necessary subjective conditions of appearances for us that they necessarily do not attach to things in themselves as they exist in the mind-independent world apart from us. In other words his argument depends, according to Hermann, on ``the erroneous assumption that already the a priori character of these notions robs them of the objective significance of determining reality in itself.'' \citep[p. 271]{hermann1935a}.

Hermann is in fact echoing similar statements by both \citet[pp. 190--197]{nelson1970} and \citet[pp. xxiv--xxv]{fries1828}. Friesians as a rule criticised Kant's arguments from `formal idealism'---a moniker which appears only in the second edition of the Critique---as well as the second edition emendations to the Critique more generally. In particular they objected to Kant's substitution of the subjective deduction of the first edition for the transcendental or `objective' deduction of the second edition, despite Kant's assertion that these are merely different ways of presenting one and the same doctrine (Bxxxviii). In light of our discussion in \S\S \ref{sec:negalt}--\ref{sec:firstprinc} it should be clear, however, that these thinkers have misinterpreted Kant's doctrine of formal idealism and the notion of objectivity appealed to therein. It should also be clear, though, that these thinkers did not interpret Kantian-Friesian doctrine ontologically themselves. They rather saw Kant as confused. Nelson writes:

\begin{quote}
[Kant's question of] how items of knowledge can relate to objects is wrongly put. The relationship of knowledge to its object is not a topic for scientific investigation at all, for it is impossible to compare knowledge with its object. In order to check on the objective validity of our knowledge, its agreement with the object, we would need to know the object independently of our knowledge. But we know it only by the knowledge which we have of it. So we should have to take this knowledge as valid, in order to compare it with the object. We should be moving in a circle. In another context Kant realized that knowledge cannot be shown to agree with its object, but in his attempt at an Objective Deduction he got involved in this fallacy. What is correct in this Deduction---and it does contain many correct and important points---is really concerned with a totally different question, not the objective validity, but the mutual relationships of items of knowledge, taken subjectively; the agreement of \emph{a priori} knowledge with the experience of which it is a condition. Take it in this way, as a question about the mutual relationship of items of knowledge, and Kant's answer is entirely correct. \citep[p. 191]{nelson1970}.
\end{quote}

Part of the misunderstanding here seems to be that Nelson (and Hermann) take Kant's use of the term `object' in a stronger sense than Kant actually intends. As we have seen, an `object' for Kant is nothing other than an \emph{objective determination} of the manifold of appearances in accordance with the synthetic a priori principles for the application of the categories, where these latter are rules for the synthesis of sensible intuition as such, whose pure forms are space and time. There is no further ontological inference here (either positive or negative) to the nature of things in themselves existing apart from us. None of this is in conflict with anything asserted by Nelson in the just quoted passage.

Regarding the `other context' mentioned by Nelson: This is most likely Kant's discussion of the antinomies. These arise, Nelson explains, ``through our silently assuming that Space and Time could contain a Whole of all existing things'' \citeyearpar[p. 170]{nelson1970}, and are resolved via the doctrine of transcendental idealism properly construed:

\begin{quote}
We can only describe the phenomena perceived as far as our experience goes, though no limits are set to the progress of that experience. So there is indeed one Nature, in space and time, one network of phenomena, bound by necessary laws. But it is not a Universe, not an absolute Whole of all existing things. Knowledge of nature is knowledge only of phenomena. \citep[p. 171]{nelson1970}.
\end{quote}

In her own essay, Hermann presents a variation of this argument that is set in a modern physical context. She first asks us to consider the principle of cause and effect, and notes that it is not really possible, strictly speaking, to apply it even within classical physics, for the relation of cause and effect presupposes that we can distinguish between two contiguous states of an object, an earlier and a later state. However the temporal evolution of classical physical states---which in the formalism of classical physics is expressed by the fact that its laws are given in the form of differential equations---is \emph{continuous}. This means, however, that for every state of the system $\omega'$ that comes earlier than $\omega$, there is actually a state $\omega''$ in between $\omega'$ and $\omega$, and thus properly speaking ``there are no temporally \emph{contiguous} states and hence for no state of a system can one specify another that has directly brought it about or has been caused by it.'' \citep[p. 272, emphasis added]{hermann1935a}.

To illustrate this, Hermann invites us to consider the `acceleration field' which surrounds the Earth in the context of Newtonian gravitational theory. That is, Newtonian theory specifies a law that associates, with a given point in the space surrounding the Earth, a particular acceleration that will be felt by any body situated at that point \citep[cf.][p. 261]{stein1967}. Thus one might say that this law causally links the presence of a body at that point (the `cause') with the acceleration imparted to it there (the `effect'). This acceleration, however, is defined via a differential equation, i.e. as the change in velocity experienced at the given point. `Change in velocity', however, is not an independent physical process but rather a ``\emph{relation} that obtains between the instantaneous velocity and the subsequent course of nature'' \citep[p. 272]{hermann1935a}, i.e. the further points on the object's trajectory.

In fact it is impossible, Hermann argues, to determine the acceleration imparted to an object existing \emph{at} a specific point in space \emph{at} the specific instant that it is there, since by considering only this single (idealised) spatiotemporal point one can in no way determine the object's velocity. Rather one must consider a wider space-time region and then compute the change in velocity of the object within that region. In considering a wider region, however, we thereby go beyond the particular point at which the instantaneous acceleration imparted to the object has been attributed. Further, within this wider region the object will experience additional changes in acceleration as it moves about. So we see that in this case it is not really possible, strictly speaking, to isolate the effect (the acceleration) at a specific point for the purposes of attributing it to a cause (i.e. that it was present at that point).

We encounter similar difficulties, according to Hermann, when we try to characterise the state description of a (classical) physical system as a fully objective description (in the strong sense) of the properties possessed by that system. For the physical quantities attributed to the system via its state description (mass density, velocity, etc.) must be specified for every point comprising its extension as well as for every point along its trajectory. But these specifications make use of differential equations which, as we just saw, actually require extended spatiotemporal regions for their determination. So we cannot say, according to Hermann, that the system `really bears' these properties which we ascribe to it via its state description:

\begin{quote}
One can thus say nothing as to the specification of the properties of physical objects or events that determine these as they are constituted in themselves; rather, the alleged properties of physical systems in truth only specify certain relations between the parts of the system, without these parts being themselves unambiguously specifiable. \citep[p. 273]{hermann1935a}.
\end{quote}

We see, therefore, that even in classical physics the conception of things which exist in space and time and which stand in causal relationships to one another is in fact of only limited applicability; we can in principle successfully describe phenomena in accordance with this conception, but only if we do so `in the limit'. That is,

\begin{quote}
As long as one is dealing only with finite, spatially and temporally extended physical systems whose own inner structure need not be considered beyond a certain limit, one escapes the difficulties of the limiting process that make impossible the full application of the concepts of cause and of substance. \citep[p. 273]{hermann1935a}.
\end{quote}

In this sense one can employ the fundamental concepts of `substance', `causality', etc. within physical theory only if one understands one's use of these concepts in this context in terms of an `analogy' \citep[p. 273]{hermann1935a}. For a complete specification of the object in spatiotemporal terms is on the basis of the above considerations impossible. But as analogies, i.e. in these concepts' application to finitely specified objects---objects for which everything beyond a certain specified bound has been `cut' away from our description of them---these fundamental concepts can be used fruitfully, and indeed are indispensable. For physical investigation always begins with the observation of a finite extended system which we conceive of as an incompletely specified thing existing in space and time in interaction with its surroundings. And it is the very fact that this specification is incomplete that presents us with the occasion for better determining the object of our investigations beyond our arbitrary initial `cut' \citep[p. 274]{hermann1935a}.\footnote{Cf. Harper's \citeyearpar[]{harper2011} account of `theory-mediated measurements'.} And indeed we do so. And yet although we can move this cut further and further forward we can never push it away entirely, for it alone guarantees that we can fruitfully describe the object of our investigation as a thing, albeit only by way of analogy; i.e. not as a thing considered in itself apart from the conditions under which it is represented by us, but always only in relation to the spatiotemporal limitation imposed upon our representation of it by the forms of our sensibility.

One might worry that this doctrine seems to depart substantially from Kant's. For Kant, recall, it is constitutive of what it means to be an object that it is fully determined in space and time in accordance with the mathematical and dynamical synthetic a priori principles. Hermann seems to be arguing, however, that in fact a full determination of this kind is impossible. But from that it would seem to follow that Kantian objects are strictly speaking impossible as well.

Now it is true, as we have seen, that for Kant the concept of an object of cognition presupposes the ascription to it of determinate spatiotemporal parameters. And it is also true that for Kant, our representations of space and time are representations of continuous, infinitely divisible, quantities (B211, B255). Yet it is not a part of Kant's doctrine that an object must be determinable beyond limit. This would only follow (as Hermann is of course aware) if we took our spatiotemporal characterisation of the object to pertain to it as it is in itself. In that case a truly objective characterisation of the object would indeed require us to describe it exhaustively via a completed infinity of spatiotemporal divisions. Kant's synthetic a priori principles, however, are only applicable to appearances, and completed infinities of this kind are not to be found in experience. To assume so can result only in antinomies (A523--527/B551--555).\footnote{See also \citet[\S 4]{bell2017}.} In this sense we can see how Hermann's discussion constitutes, exactly as she maintains, an argument for transcendental idealism motivated by a consideration of the antinomies.

Now Hermann points out that one might respond to this line of argument as follows \citep[p. 275]{hermann1935a}: One might concede that the conception of things existing objectively (in the strong sense) in space and time must dissolve, if pushed to its inevitable logical conclusion, into a network of spatiotemporal relations which \emph{cannot} be characterised in terms of such things, and therefore that such a conception is impossible. Nevertheless one might maintain that it is possible that further investigation into the natural world will reveal that this relational network itself is objectively determinable via a consideration of things (in the weaker sense) describable in space and time. In other words it could be that the spatiotemporal relational network is itself objective even if it cannot literally be construed as pertaining to objects (thought of in the strong sense).\footnote{It is possible that the view represented by the hypothetical objector here may actually be that of Nelson. In the context of practical philosophy, for instance, Nelson maintained that the moral law, although only discoverable empirically and `subjectively', was nevertheless objective. This was a conclusion with which Hermann disagreed \citep{leal2017}.}

Far from disagreeing, Hermann maintains that this is a question which can only be decided by empirical research; and it has in fact, she argues, actually been answered in the negative by quantum theory. Quantum theory shows us that this relational network is not objective but rather `split'.\footnote{See also \citet[]{crull2017}.} \citet[p. 275]{hermann1935a} writes:

\begin{quote}
[Quantum mechanics] does away with the notion that these relational networks should be determined at any rate through objective circumstances of things in space and time, and shows them in turn to depend on the manner in which the observer obtains knowledge of the system. \citep[p. 275]{hermann1935a}.
\end{quote}

This is so because, as we have seen, our ascription of a causal history to a quantum object depends on the experimental context under which that object is considered. In the experiment (described earlier) which we set up to determine the momentum of a given quantum system of interest, for example, we explain the fact that the pointer took on a particular reading by positing that the quantum system had a particular momentum at the time of its interaction with the diaphragm. That is, this posit, in conjunction with the classical conservation theorems, accounts for the fact that the pointer reading took on a particular value in the context of this measurement. In performing such a measurement, however, we cut ourselves off from ascribing a determinate value of position to the quantum system of interest at the time of its interaction with the diaphragm, for the determinateness of our attribution of position to the quantum system varies inversely with the determinateness of our attribution of momentum. Further, not only does what we learn about the quantum system depend on our observational context, it is also the case that (see also \S \ref{sec:conceptind} above) these observational contexts are in general incompatible. Yet as Hermann points out, ``the limit up to which one or the other model finds application is \emph{not itself an objective property of the object}'' \citep[p. 270, emphasis added]{hermann1935a}.

Within quantum mechanics there is no viewpoint, therefore, from which one can give the quantum object a complete characterisation in the classical sense of determining both its position and its momentum; it is therefore not an object in that sense. Nevertheless it is possible to speak objectively about the quantum object \emph{from within} a particular observational context. We can say, that is, that (experimental error notwithstanding) the fact that the quantum object had a particular momentum was \emph{the cause} of the pointer's having indicated a particular value, because there is no information which we lack in order to complete our characterisation of the object from the perspective of that experimental context---in other words we have a complete characterisation of `the object' that our experiment was set up to ascertain: the photon's momentum.

To come back now to the question we first considered in \S \ref{sec:hermann}: Is Hermann's view actually Kantian? In fact there is a sense in which it is Kantian and a sense in which it is not. Let us begin with the latter: Kant has developed his system a priori, beginning from a posit regarding the nature of our cognition as involving the contributions of both the faculties of sensibility and understanding, and has provided an account of the a priori forms of these faculties (space and time and the categories). From this he has deduced a number of synthetic a priori principles which are presupposed by the concept of objective determination as such. In the context of physical theory these principles imply that the objective determination of something must be such as to ascribe to it determinate values for both its position and its momentum. \emph{Pace} Hermann, the one-sided `objects' described in the previous paragraph do not satisfy this criterion, however, and in this sense they are not objects of theoretical cognition to which the principle of causality could be applicable.

In a different sense, however, Hermann's viewpoint is indeed quite Kantian. For after all, Kantian philosophy teaches us that there are multiple ways of considering phenomena. We may consider ourselves as either objects of theoretical or practical (i.e. ethical) inquiry, for instance, and each of these domains of inquiry comes with its own criteria for cognition. Moreover since our cognition of objects is limited to appearances it does not follow that any one domain can be thought of as either ontologically superior or reducible to the other.\footnote{Cf. \citet[p. 48]{allison2004}, \citet[p. 247ff]{nelson1971}.} Further, within each domain distinct notions of justification are operative, along with distinct notions of causality---most generally thought of as change in accordance with law \citepalias[4:446]{kant1785}---and the other fundamental concepts of the Kantian system. ``The will is a kind of causality'', Kant writes, ``belonging to living beings insofar as they are rational'' \citepalias[4:446]{kant1785}.

What quantum mechanics teaches us, according to Hermann, is that even \emph{within} a single domain of inquiry, such as physical science, one can be confronted with a splitting of perspectives in the above sense:

\begin{quote}
the splitting of truth goes deeper than philosophy and natural science had previously assumed. It penetrates into the physical knowledge of nature itself; instead of merely delimiting its scope against other possibilities for grasping reality, it separates various equally legitimate representations within the physical description that cannot be unified into a single picture of nature \citep[p. 277]{hermann1935a}.
\end{quote}

Beginning from such motivations, therefore, it seems natural for a Kantian to want to relativise the concept of causality to different perspectives in this way within the domain of physical science. And yet for an orthodox Kantian this way seems barred for the reasons which we discussed above. Hermann, however, is not an orthodox Kantian but a Friesian neo-Kantian. And as we saw in \S\ref{sec:fries-nelson}, for Friesians the method of critical philosophy is not a priori but \emph{empirical}. Friesians are allowed to learn and adapt the critical philosophy in the face of the challenges posed to it by quantum mechanics in a way that orthodox Kantians seemingly cannot. The way is open, therefore, for Hermann's relativisation of the concept of causality (and along with it, objectivity) within physical science. And she need not deny the central tenets of critical philosophy as it was understood by the Friesian school of neo-Kantianism---namely its core conception of the critical \emph{method}---in order to do so.

\section{}
\label{sec:conc}

Let us now summarise our discussion and conclude. I introduced and motivated Kant's critical philosophy---including his doctrine of transcendental idealism and the intellectual route by which he arrived at it---in \S\S\ref{sec:negalt}--\S\ref{sec:firstprinc}, and then his principles for synthetic a priori cognition in \S\ref{sec:synthap}. In \S\ref{sec:quantmech} I considered the challenges for Kant's view which stem from the emergence of quantum theory and in \S\S\ref{sec:conceptind}--\ref{sec:schlick} I showed how these principles may be maintained on an orthodox (i.e. discursive) interpretation of Kantian philosophy along Bohrian lines. In \S\S\ref{sec:hermann}--\ref{sec:herm-transcideal} I then introduced the views of the Friesian neo-Kantian philosopher Grete Hermann and in particular her understanding of the general situation regarding Kantian critical philosophy \emph{vis \'a vis} quantum theory.

I argued that on an orthodox Kantian viewpoint---for which the view of Niels Bohr may be taken as representative (though he never called himself such)---the point emphasised by quantum mechanics is that our fundamental (and necessary) forms of knowledge are ultimately idealisations and as such are applicable only within a limited domain. And yet we may nevertheless---because of this very limitation---continue to employ these idealisations outside of this domain in order to convey information to one another in a qualified way; i.e. with the understanding that the information conveyed in this way does not in general constitute a standpoint-independent description of the phenomenon under investigation. As for Hermann---whom we may take to represent the Friesian viewpoint---I showed that there is, in addition to this, a deeper lesson as well: Quantum mechanics teaches us that physical knowledge is fundamentally `split'; for Hermann, that is, there is a notion of objectivity that is applicable to quantum phenomena, however it is a form of objectivity that is intrinsically perspective-dependent.

Which of these positions is to be preferred? I will not attempt to answer this question definitively here. There were other neo-Kantian thinkers---to name one: Ernst Cassirer \citeyearpar[]{cassirer1936}---who commented on quantum mechanics, and any judgement regarding the correct path for a Kantian to take in the face of quantum theory should more properly attend upon a detailed consideration of these others in addition to the two thinkers which we have considered here. But perhaps we may make one or two tentative statements in the meantime. A merit of Bohr's view is that it emphasises the lesson of transcendental idealism, i.e. that the epistemic primacy of the conditions under which we can assert that we have cognised some object does not imply the ontological primacy of those conditions. Quantum mechanics comes out as a striking example of this on the Bohrian view. For Hermann, on the other hand, the lesson is that in general objectivity (in the weak sense) must be thought of as relativised within physical science to a particular perspective---i.e. a particular measurement context---within which we may continue to meaningfully employ principles such as that of causality to the representations of phenomena that are confined to that perspective. Quantum mechanics teaches us this, on Hermann's view. And it is a merit of her viewpoint that it makes sense of the fact that quantum mechanics can teach us \emph{anything at all} about the a priori forms of our cognition. Moreover it hints, perhaps, at a more general account of the way that the a priori forms of our cognition---what in another context have been called `framework assumptions'---may change their scope and meaning over time.

\bibliographystyle{apa-good}
\bibliography{Bibliography}{}

\end{document}